\documentclass[11pt]{article}
\usepackage{epsfig}
\newcommand{\la}[1]{\label{#1}}
\newcommand{\be}{\begin{equation}}
\newcommand{\ee}{\end{equation}}
\newcommand{\ba}{\begin{eqnarray}}
\newcommand{\ea}{\end{eqnarray}}
\newcommand{\rmi}[1]{{\mbox{\scriptsize #1}}}
\newcommand{\fig}{Fig.~}
\newcommand{\eq}{Eq.~}

\newcommand{\eqs}{Eqs.~}
\newcommand{\nr}[1]{(\ref{#1})}
\newcommand{\tr}{{\rm Tr\,}}

\newcommand{\fr}[2]{{\frac{#1}{#2}\,}}
\newcommand{\msbar}{{\overline{\mbox{\rm MS}}}}

\newcommand{\e}{\epsilon}

\newcommand{\bG}{{\beta_G}}

\newcommand{\Nc}{N_{\rm c}}

\newcommand{\CA}{\Nc}

\renewcommand{\bG}{B_\rmi{G}}    

\def\lsi{\raise0.3ex\hbox{$<$\kern-0.75em\raise-1.1ex\hbox{$\sim$}}}
\def\gsi{\raise0.3ex\hbox{$>$\kern-0.75em\raise-1.1ex\hbox{$\sim$}}}
\newcommand{\lsim}{\mathop{\lsi}}

\newcommand{\bmu}{\bar\mu}
\newcommand{\tinymsbar}{{\overline{\mbox{\tiny\rm{MS}}}}}

\makeatletter \@addtoreset{equation}{section} \makeatother

\makeatletter
\renewcommand\section{\@startsection {section}{1}{\z@}%
                                   {-5.5ex \@plus -1ex \@minus -.2ex}
                                   {2.3ex \@plus.2ex}%
                                   {\normalfont\large\bfseries}}
\renewcommand\subsection{\@startsection{subsection}{2}{\z@}%
                                     {-3.25ex\@plus -1ex \@minus -.2ex}%
                                     {1.5ex \@plus .2ex}%
                                     {\normalfont\normalsize\bfseries}}
\renewcommand\thesection {\@arabic\c@section}
\renewcommand\thesubsection   {\thesection.\@arabic\c@subsection}
\renewcommand{\@seccntformat}[1]{%
\csname the#1\endcsname.\hspace{1.0em}}
\makeatother
\begin{document}

\begin{titlepage}
\begin{flushright}
BI-TP 2004/36\\
CERN-PH-TH/2004-236\\
HIP-2004-48/TH\\
hep-lat/0412008\\
\end{flushright}
\begin{centering}
\vfill
 
{\Large{\bf Plaquette expectation value and gluon condensate}}

\vspace{0.2cm}
 
{\Large{\bf in three dimensions}}

\vspace{0.8cm}

A. Hietanen$^{\rm a}$, 
K. Kajantie$^{\rm a}$, 
M.~Laine$^{\rm b}$, 
K.~Rummukainen$^{\rm c,d,e}$, 
Y. Schr\"oder$^{\rm b}$ 

\vspace{0.8cm}

{\em $^{\rm a}$%
Theoretical Physics Division, 
Department of Physical Sciences, \\ 
P.O.Box 64, FI-00014 University of Helsinki, Finland\\}

\vspace{0.3cm}

{\em $^{\rm b}$%
Faculty of Physics, University of Bielefeld, 
D-33501 Bielefeld, Germany\\}

\vspace{0.3cm}

{\em $^{\rm c}$%
Department of Physics, Theory Division, CERN, CH-1211 Geneva 23,
Switzerland\\}

\vspace{0.3cm}

{\em $^{\rm d}$%
Department of Physics, University of Oulu, 
P.O.Box 3000, FI-90014 Oulu, Finland\\}

\vspace{0.3cm}

{\em $^{\rm e}$%
Helsinki Institute of Physics,
P.O.Box 64, FI-00014 University of Helsinki, Finland\\}

\vspace*{0.8cm}
 
\mbox{\bf Abstract}

\end{centering}

\vspace*{0.3cm}
 
\noindent
In three dimensions, the gluon condensate of pure SU(3) gauge theory
has ultraviolet divergences up to 4-loop level only. By subtracting 
the corresponding terms from lattice measurements of the plaquette 
expectation value and extrapolating to the continuum limit, we extract
the finite part of the gluon condensate in lattice regularization. 
Through a change of regularization scheme to $\msbar$ and (inverse)
dimensional reduction, this result would determine the first non-perturbative 
coefficient in the weak-coupling expansion of hot QCD pressure.
\vfill
\noindent
 

\vspace*{1cm}
 
\noindent
December 2004

\vfill

\end{titlepage}

%
\section{Introduction}
\la{se:introduction}

As non-Abelian gauge theories in three and four dimensions are
confining, their properties need, in general, to be studied
non-perturbatively.  If the observables considered involve momenta 
or masses ($M$) large compared with the confinement scale, however, 
then the conceptual framework of the 
operator product expansion~\cite{ope} may allow to isolate the
non-perturbative dynamics into only a few (gluon) condensates, while
the rest of the answer can be computed by perturbative means.  
A classic example is the mass of a heavy quark--anti-quark bound
state~\cite{lv}.  The task faced by numerical lattice 
simulations might then get
significantly simplified, as local condensates are simpler to
measure with controlled systematic errors
than correlation functions of heavy states.

On the other hand, the physical interpretation of a ``bare'' lattice
measurement of a gluon condensate is non-trivial. The reason is 
that the condensate is represented by the expectation value of 
a dimensionful singlet operator and, in general, contains
ultraviolet divergences of the same degree as its dimension. 
Operator product expansion type relations are
often derived employing dimensional regularization, since the system then
only contains one large parameter ($M$) rather than two ($M$ and the
momentum cutoff), which simplifies the derivation
considerably. Making use of lattice results
in such a context requires then a
transformation from lattice to continuum
regularization. While in principle a well-defined 
perturbative problem (see, e.g.,
Refs.~\cite{hh}), this is in practice somewhat problematic in four
dimensions, given that there are contributions from all
orders in the loop expansion.

The observable we consider in this paper is the (lowest-dimensional)
singlet gluon condensate in three dimensions (3d), measured with pure SU(3)
gauge theory.  As 3d pure Yang-Mills theory is super-renormalisable,
the problem of changing the regularization scheme becomes solvable:
there are ultraviolet divergences up to 4-loop level
only~\cite{framework}. Furthermore, as we will elaborate in the
following, all the divergences have been determined recently. 
These advances allow us to obtain a finite ``subtracted'' continuum value 
for the gluon condensate in lattice regularization. A conversion
to the $\msbar$ scheme, amounting to the (perturbative) computation of the 
constant $c_4'$ in \eq\nr{eq:betaG} below, 
remains however a future challenge.

There might be various physics settings where the 3d gluon condensate
finds practical applications.  The one that motivated us, is that this
condensate appears in high-temperature physics, where the temperature
$T$ plays the role of the mass scale $M$ mentioned above. Indeed 3d
pure Yang-Mills theory determines the leading non-perturbative
contribution to the weak-coupling expansion of the pressure 
(and a number of other quantities) of physical 
QCD~\cite{linde,gpy}, through a conceptual counterpart of the
operator product expansion, called finite-temperature dimensional
reduction~\cite{dr,generic,bn}. Other applications might exist as well.

The plan of this paper is the following.
In Sec.~\ref{se:theory}, we specify the observables considered and 
discuss the theoretical setting of our study. Numerical results from 
lattice Monte Carlo simulations are reported in Sec.~\ref{se:measurements}, 
and we conclude in Sec.~\ref{se:conclusions}.

%
\section{Theoretical setting}
\la{se:theory}

We start this Section by formulating 
the observables that we are interested in, 
in the formal continuum limit of the theory. The ultraviolet (UV) divergences
appearing in loop contributions are at this stage regulated through the use of 
dimensional regularization. Later on we go over to lattice 
regularization, in order to give a precise non-perturbative meaning to the  
observables introduced, allowing for their numerical determination.
 
The Euclidean continuum action of pure SU($\Nc$) 
Yang-Mills theory can be written as 
\ba 
 S_\rmi{E} =  \int \! {\rm d}^d x\, {\cal L}_\rmi{E}
 \;, \qquad \la{eq:SE} 
 {\cal L}_\rmi{E} = \fr1{2g_3^2} \tr [F_{kl}^2 ]
 \;. \la{eq:LE}
\ea
Here $d=3-2\epsilon$, 
$g_3^2$ is the gauge coupling,  
$k,l=1,...,d$,  
$F_{kl} = i [D_k,D_l]$,  
$D_k = \partial_k - i A_k$, $A_k = A_k^aT^a$, 
$T^a$ are the Hermitean generators of SU($\Nc$), 
normalised as $\tr[T^a T^b] = \delta^{ab}/2$,
and repeated indices are assumed to be summed over.
Leaving out for brevity gauge fixing and Faddeev-Popov terms, 
the  ``vacuum energy density'' reads
\be
 f_\tinymsbar \equiv - \lim_{V \to \infty} \frac{1}{V} 
 \ln \biggl[ \int \! {\cal D} A_k 
 \, \exp\Bigl( -S_\rmi{E} \Bigr) \biggr]_\tinymsbar
 \;, \la{eq:f}
\ee
where $V$ is the $d$-dimensional volume, 
${\cal D} A_k$ a suitable (gauge-invariant) functional integration measure, 
and we have assumed the use 
of the $\msbar$ dimensional regularization scheme to remove
any $1/\epsilon$ poles from the expression. 
We note that $f_\tinymsbar$ has the dimensionality [GeV]$^d$.

In strict dimensional regularization, $f_\tinymsbar$ of course vanishes 
order by order in the loop expansion, due to the absence of any mass scales 
in the propagators. This behaviour is unphysical, however, 
and due to an exact cancellation
between UV and infrared (IR) divergences; for an explicit 
discussion at 3-loop level in a related case, 
see Appendix B of Ref.~\cite{aminusb}. In fact
non-perturbatively the structure of $f_\tinymsbar$ 
is rather
\be
 f_\tinymsbar = -g_3^6 
 \frac{d_A \CA^3}{(4\pi)^4} 
 \biggl[
  \biggl( \frac{43}{12} - \frac{157}{768} \pi^2 \biggr) 
 \ln\frac{\bmu}{2 \CA g_3^2} + \bG 
 + \mathcal{O}(\epsilon)
 \biggr]
 \;, \la{eq:structure}
\ee
where 
$d_A \equiv \Nc^2-1$, and we 
have introduced an $\msbar$ scheme scale parameter $\bmu$.
The coefficient of the logarithm in \eq\nr{eq:structure} has been 
determined in Ref.~\cite{sun} with a perturbative 4-loop computation,  
by regulating all the propagators by a small mass 
scale $m_\rmi{G}$, and sending $m_\rmi{G}\to 0$ only after 
the computation (see also Ref.~\cite{gsixg}). 
The non-perturbative constant part $\bG$,\footnote{%
  In Ref.~\cite{gsixg}, $\bG$ was denoted by $\beta_\rmi{G}$, 
  but we prefer to introduce a new notation here, in order to avoid
  confusion with the coupling constant $\beta$ appearing 
  in~\eq\nr{eq:Sa}. The subscript G might refer to gluons.
  }  
which actually 
is a function of $\Nc$, is what we would ultimately like to determine. 

One direct physical application of $\bG$ is that it 
determines the first non-perturbative contribution to the weak-coupling
expansion of the pressure $p$ 
of QCD at high temperatures~\cite{linde,gpy}. 
To be precise, this contribution 
is of the form $\delta p = d_A \CA^3 g^6 T^4 \bG/(4\pi)^4$,
where $g^2$ is the renormalised QCD gauge coupling.  Terms up to 
order $\mathcal{O}(g^6\ln(1/g))$ are, in contrast, perturbative, and 
all known by now~\cite{gsixg}.

For future reference, we note that given $f_\tinymsbar$, 
we immediately obtain the gluon condensate: 
\ba
 \frac{1}{2 g_3^2} \Bigl\langle
 \tr [F_{kl}^2] 
 \Bigr\rangle_\tinymsbar & \equiv &  
 - g_3^2 \frac{\partial}{\partial g_3^2} f_\tinymsbar 
 \\ & = &  
 3 g_3^6 
 \frac{d_A \CA^3}{(4\pi)^4} 
 \biggl[
  \biggl( \frac{43}{12} - \frac{157}{768} \pi^2 \biggr) 
 \biggl( 
 \ln\frac{\bmu}{2 \CA g_3^2} -\fr13
 \biggr)
 + \bG 
 + \mathcal{O}(\epsilon)
 \biggr]
 \;.
 \la{eq:condensate}
\ea

We now go to the lattice. In the standard Wilson discretization,  
the lattice action, $S_a$, corresponding to \eq\nr{eq:SE}, reads 
\ba
 S_a & = & 
 \beta \sum_{\bf x} \sum_{k < l}
 \Bigl( 1 - \frac{1}{\Nc} \mathop{\mbox{Re}} \tr [P_{kl}({\bf x}) ] \Bigr)
 \;, \la{eq:Sa}
\ea
where   
$P_{kl}({\bf x})=U_k({\bf x})U_l({\bf x}+k)U_k^{-1}({\bf x}+l)U_l^{-1}({\bf x})$
is the plaquette,
$U_k({\bf x})$ is a link matrix, 
${\bf x}+k\equiv  {\bf x}+a\hat\e_k$, where 
$a$ is the lattice spacing and $\hat\e_k$ 
is a unit vector,  and 
\be
 \beta \equiv \frac{2 \Nc}{g_3^2 a}
 \;. \la{eq:beta}
\ee
Note that the gauge coupling does not get renormalised in 3d, 
and the parameters $g_3^2$ appearing in \eqs\nr{eq:SE}, \nr{eq:beta} 
can hence be assumed finite and equivalent. 
The observable we consider is still the vacuum energy density, \eq\nr{eq:f}, 
which in lattice regularization reads
\be
 f_a \equiv - \lim_{V \to \infty} \frac{1}{V} 
 \ln \biggl[ \int \! {\cal D} U_k 
 \, \exp\Bigl( -S_a \Bigr) \biggr]
 \;, \la{eq:fa}
\ee
where ${\cal D} U_k$ denotes integration over link matrices
with the gauge-invariant Haar measure.

Now, being in principle
physical quantities, the values of $f_\tinymsbar$ and $f_a$ must agree, 
provided that suitable vacuum counterterms are added to the theory.
Due to super-renormalizability, there can be such counterterms up to 4-loop
level only~\cite{framework}, and correspondingly
\ba
 \Delta f & \equiv & 
 f_a - 
 f_\tinymsbar
 \\ & = & 
 C_{1} \, \frac{1}{a^3} \biggl( \ln\frac{1}{a g_3^2} + C_1' \biggr )
 + 
 C_{2} \, \frac{g_3^2}{a^2} 
 + 
 C_{3} \, \frac{g_3^4}{a}
 + 
 C_{4} \, g_3^6 \biggl( \ln\frac{1}{a\bmu} + C_4' \biggr)
 + \mathcal{O}(g_3^8 a)
 \;, \la{eq:Deltaf}
\ea 
where the $C_i$ are dimensionless functions of $\Nc$. 
The values of $C_1,C_2,C_3,C_4$
are known, as we will recall presently; 
$C_1'$ is related to the precise normalisation
of the Haar integration measure and 
void of physical significance; and $C_4'$ is 
unknown as of today. 

Correspondingly, the gluon condensates, i.e. the 
logarithmic derivatives of $f_\tinymsbar$, $f_a$
with respect to $g_3^2$, 
can also be related by 
a perturbative 4-loop computation. Noting that three-dimensional rotational 
and translational symmetries and the reality of $S_a$
allow us to write 
\be
 - g_3^2 \frac{\partial}{\partial g_3^2} 
 f_a
 = \frac{3\beta}{a^3}
 \Bigl\langle 1 - \frac{1}{\CA} \tr[P_{12}] \Bigr\rangle_a
 \;, \la{eq:plaq}
\ee  
and employing \eqs\nr{eq:condensate}, \nr{eq:Deltaf}, we obtain finally 
the master relation
\be
 8 \frac{d_A\CA^6}{(4\pi)^4} \bG =
 \lim_{\beta\to\infty} \beta^4
 \biggl\{
 \Bigl\langle
 1 - \frac{1}{\CA} \tr[P_{12}]
 \Bigr\rangle_a - 
 \biggl[ 
 \frac{c_1}{\beta} + \frac{c_2}{\beta^2} + 
 \frac{c_3}{\beta^3} + \frac{c_4}{\beta^4}
 \Bigl(
 \ln\beta + c_4' 
 \Bigr)
 \biggr]
 \biggr\} 
 \;. \la{eq:betaG}
\ee
The values of the constants $c_1,...,c_4'$ are trivially
related to those of $C_1,...,C_4'$ in \eq\nr{eq:Deltaf}. 

Now, a straightforward 1-loop computation yields
\be
 c_1 = \frac{d_A}{3} \approx 2.66666667 
 \;, \la{eq:c1}
\ee
where the numerical value applies for $\Nc = 3$. 

The 2-loop term is already non-trivial: it 
was first computed in four dimensions in Ref.~\cite{gr},
and in three dimensions
in Ref.~\cite{hk}. The 3d result can be written in the form
\be
 c_2 = -\fr23 \frac{d_A \CA^2}{(4\pi)^2}
 \biggl( 
 \frac{4\pi^2}{3 \CA^2} + \frac{\Sigma^2}{4} - \pi \Sigma 
 - \frac{\pi^2}{2} + 4 \kappa_1 + \fr23 \kappa_5
 \biggr) = 1.951315(2)
 \;, \la{eq:c2}
\ee
where the coefficients $\Sigma$, $\kappa_1$ can be 
found in Refs.~\cite{framework,contlatt}, and we have defined
\be
 \kappa_5 =
 \frac{1}{\pi^4}\int_{-\pi/2}^{\pi/2}\!\! {\rm d}^3x\, {\rm d}^3y
 \frac{\sum_i {\sin}^2x_i {\sin}^2(x_i+y_i) {\sin}^2y_i }
 {\sum_i{\sin}^2x_i \sum_i{\sin}^2(x_i+y_i) \sum_i{\sin}^2y_i}
 =1.013041(1)
 \la{eq:kappa5} \;.
\ee
The numbers in parentheses in \eqs\nr{eq:c2}, \nr{eq:kappa5}
indicate the uncertainties of the last digits.

The 3-loop term is well known in 
four dimensions since a long time ago~\cite{acfp}, but the same 
computation has been carried out in 
three dimensions only very recently~\cite{pt}: 
\be
 c_3 = 6.8612(2)
 \;. \la{eq:c3} 
\ee
This value improves on a previous estimate
$c_3 = 6.90 \raise-0.3ex\hbox{$\stackrel{\scriptscriptstyle{+0.02}}
{\scriptscriptstyle{-0.12}}$}$~\cite{dir}, obtained through
the evaluation of the 3-loop graphs with the method of stochastic 
perturbation theory~\cite{romm}.

The value of $c_4$ follows by a comparison of \eqs\nr{eq:structure} and 
\nr{eq:Deltaf}: there is no $\bmu$-dependence in $f_a$, so that the one in 
$f_\tinymsbar$ determines the coefficient of the logarithm in $\Delta f$.
Consequently, 
\be
 c_4 = 
 8 \frac{d_A\CA^6}{(4\pi)^4}
 \biggl(
 \frac{43}{12} - \frac{157}{768} \pi^2  
 \biggr) \approx 2.92942132
 \;. \la{eq:c4} 
\ee

The knowledge of $c_1,c_2,c_3,c_4$ allows us to subtract all the divergent
contributions from the gluon condensate. A finite 4-loop term, 
parametrised by $c_4'$
in \eq\nr{eq:betaG}, however still remains. It could in principle be determined
by extending either the method 
of Ref.~\cite{pt} or of Ref.~\cite{dir} to 
4-loop level. There is the additional complication, though, that intermediate 
steps of the computation require the use of an IR cutoff, 
which then cancels once the lattice and $\msbar$ results are subtracted, in  
\eq\nr{eq:Deltaf}. This computation has not been carried out yet, and 
therefore we will not be able to determine $\bG$ in this paper. 
We can determine, however, the non-perturbative input needed 
for it (cf.\ \eq\nr{eq:final} below), the purely perturbative determination 
of $c_4'$ then remaining a future challenge.

%
\section{Lattice measurements}
\la{se:measurements}

The goal of the numerical study is to measure the plaquette
expectation value, $\langle 1 - \frac{1}{3} \tr[P_{12}] \rangle_a$, as
a function of $\beta$, such that the extrapolation in~\eq\nr{eq:betaG}
can be carried out. For each $\beta$, the infinite-volume limit needs
to be taken. Given that the theory has a mass gap, we expect that
finite-volume effects are exponentially small, if the length of the
box $L$ is large compared with the confinement scale, $\sim 1/ g_3^2$.
Writing $L = N a$, where $N$ is the number of lattice sites, 
the requirement $L \gg 1/g_3^2$ converts to $\beta/N \ll 6$
(cf.\ \eq\nr{eq:c4}). Detailed studies with other observables 
show that in practice the finite-volume effects are 
invisible as soon as $\beta/N < 1$~\cite{finitevol}. The
values of $\beta$ and $N$ that we have employed are shown in Table~\ref{table}.
Earlier lattice measurements of the same observable were carried
out with a volume $N^3 = 32^3$, with values of $\beta$ up to 
$\beta=30$~\cite{klpr}.

\begin{table}[t]

\begin{center}
\begin{tabular}{ll}
\hline\hline
$\beta$ & volumes \\
\hline\hline
12 ~~~~ & $24^3,~32^3,~48^3$ \\
16 ~~~~ & $24^3,~32^3,~48^3,~64^3$ \\
20 ~~~~ & $24^3,~32^3,~48^3$ \\
24 ~~~~ & $\{12^3,~14^3,~16^3,~20^3,~24^3\},~32^3,~48^3,~64^3$ \\
32 ~~~~ & $\{14^3,~16^3,~20^3,~24^3,~32^3\},~48^3,~64^3,~96^3$ \\
40 ~~~~ & $\{32^3\},~48^3,~64^3,~96^3$ \\
50 ~~~~ & $\{20^3,~24^3,~26^3,~28^3,~32^3,~48^3\},~64^3,~96^3,~128^3,~320^3$ \\
64 ~~~~ & $\{48^3,~64^3\},~96^3,~128^3,~320^3$ \\
80 ~~~~ & $\{64^3\},~128^3,~320^3$ \\
$[$100 ~~~~ & $128^3,~192^3,~320^3]$ \\
$[$140 ~~~~ & $\{128^3\},~192^3,~320^3]$ \\
$[$180 ~~~~ & $\{192^3\},~320^3]$ \\
\hline\hline
\end{tabular}
\end{center}

\caption[a]{The lattice spacings (parametrised by $\beta$, 
cf.\ \eq\nr{eq:beta}) and the volumes
(in lattice units, $N^3$, so that $V = N^3 a^3$) studied.
On each lattice we have collected $\sim 10^4 ... 10^6$
independent measurements. 
The lattices in curly brackets have been left out from the 
infinite-volume extrapolations, while for the lattices in square 
brackets the significance loss due to the ultraviolet subtractions 
in \eq\nr{eq:betaG}
is so large (six orders of magnitude or more) that the subtracted values 
have little effect on our final fit (see below).}

\la{table}
\end{table}

It is important to stress that the subtractions in~\eq\nr{eq:betaG} lead
to a major significance loss. Essentially, we need to evaluate numerically
the fourth derivative with respect to $\beta^{-1}$ of the 
function $\langle 1 - \frac{1}{3} \tr[P_{12}] \rangle_a$, at the point 
$\beta^{-1} = 0$. Another way to express the problem is that as the numbers
$c_1,...,c_4$ are of order unity (cf.\ \eqs\nr{eq:c1}--\nr{eq:c4}), the 
dominant term, $c_1/\beta$, is about six orders of magnitude larger than 
the effect we are interested in, $\sim 1/\beta^4$, if $\beta \sim 100$.
Therefore the relative error of our lattice measurements should be
smaller than one part in a million. We also need to know the 
coefficients $c_i$  with good precision.

Lattice-measured values of 
$\langle 1 - \frac{1}{3} \tr[P_{12}] \rangle_a$ 
are shown in \fig\ref{fig:bareplaquette}, as a function of $1/\beta$. 
In order to demonstrate the accuracy requirements we are faced with, 
\fig\ref{fig:orders} shows
$\beta^4\langle 1 - \frac{1}{3} \tr[P_{12}] \rangle_a$, before and after the 
various subtractions. It is observed from \fig\ref{fig:orders}
that after all the subtractions, 
this function indeed appears to have a finite 
limit for $\beta\to\infty$, or $1/\beta \to 0$.

\begin{figure}[t]


\centerline{
~~~\epsfysize=7.0cm\epsfbox{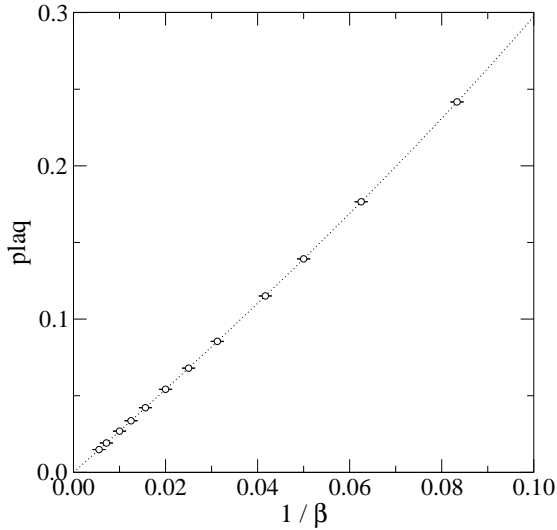}%
}


\caption[a]{The plaquette expectation value, 
``plaq'' $\equiv \langle 1 - \frac{1}{3} \tr[P_{12}] \rangle_a$,
as a function of $1/\beta$. 
Statistical errors are (much) smaller than the symbol sizes.
The dotted curve contains the four known terms 
$c_1/\beta+c_2/\beta^2+c_3/\beta^3+c_4 \ln\beta/\beta^4$ from 
\eq\nr{eq:betaG}, together with terms of the type
$1/\beta^4$, $1/\beta^5$ and $1/\beta^6$ with fitted coefficients.}

\la{fig:bareplaquette}
\end{figure}

\begin{figure}[t]


\centerline{
~~~\epsfysize=7.0cm\epsfbox{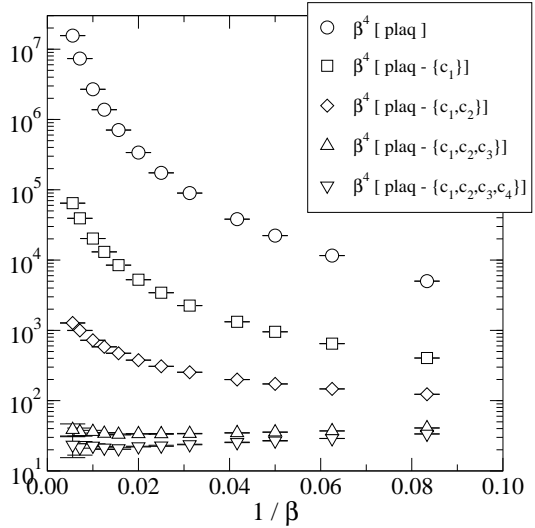}%
 }


\caption[a]{The significance loss due to the subtractions of the various
ultraviolet divergent contributions in the gluon condensate. Here again
``plaq'' $\equiv \langle 1 - \frac{1}{3} \tr[P_{12}] \rangle_a$, and the
symbols $c_i$ in the curly brackets indicate which subtractions 
of~\eq\nr{eq:betaG} have been taken into account.}

\la{fig:orders}
\end{figure}

For each $\beta$, we have carried out simulations at a number of different
lattice extents $N$; examples are shown in \fig\ref{fig:infvol}.
No significant volume dependence is observed for $\beta/N < 1$, and 
we thus estimate the infinite-volume limit by fitting a constant
to data in this range. 

\begin{figure}[t]



\centerline{
\epsfysize=5.0cm\epsfbox{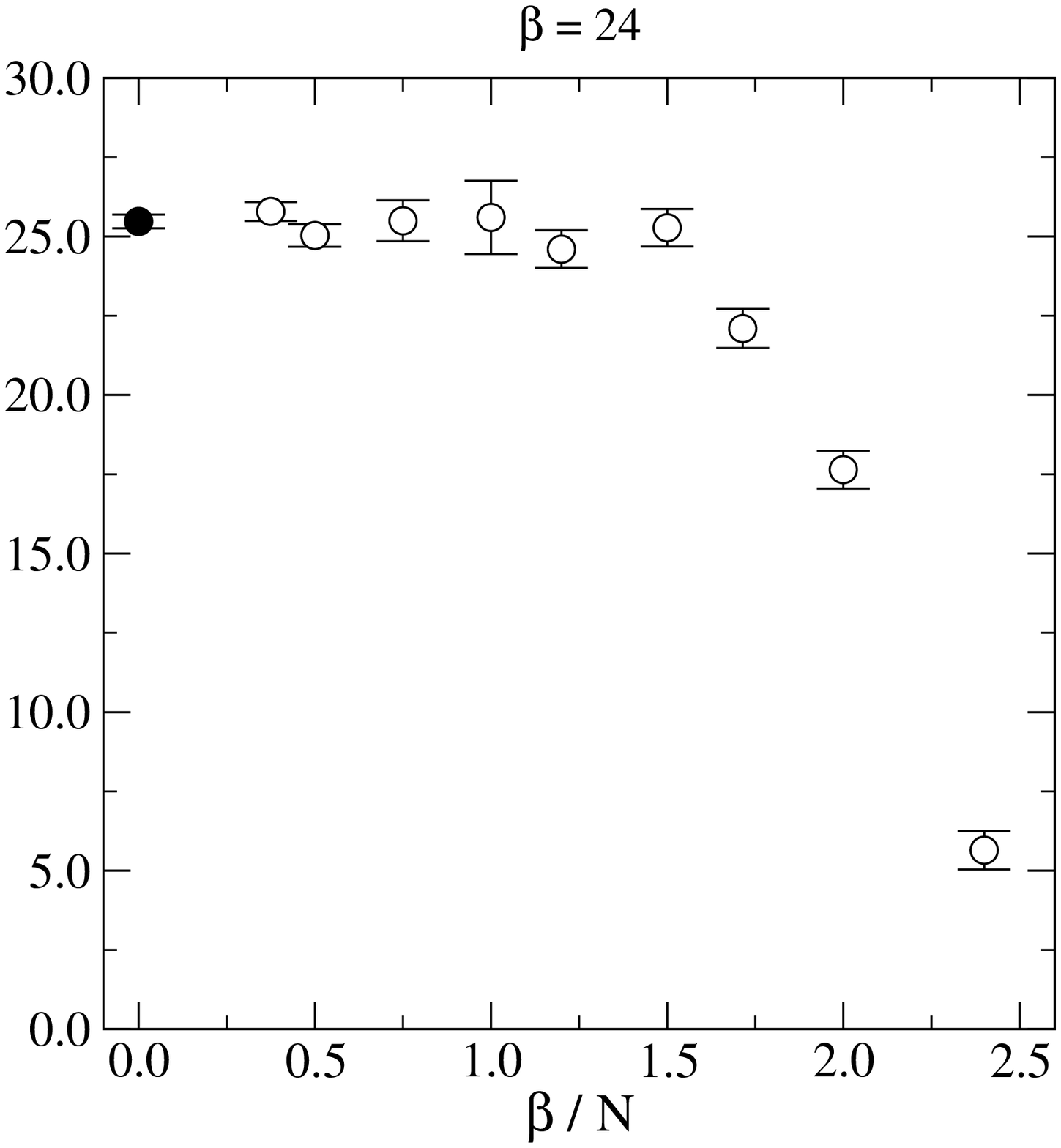}%
~~~\epsfysize=5.0cm\epsfbox{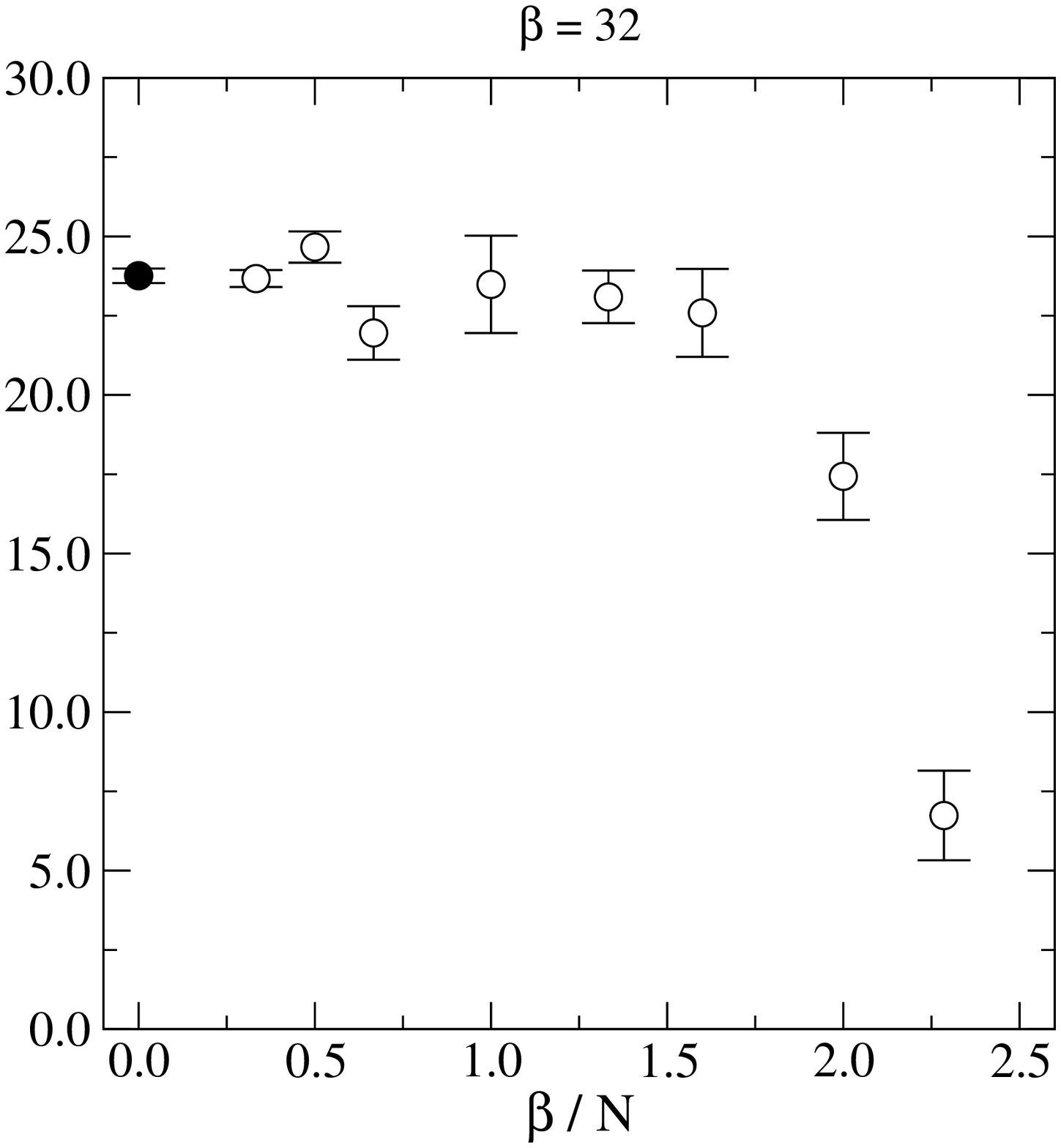}%
~~~\epsfysize=5.0cm\epsfbox{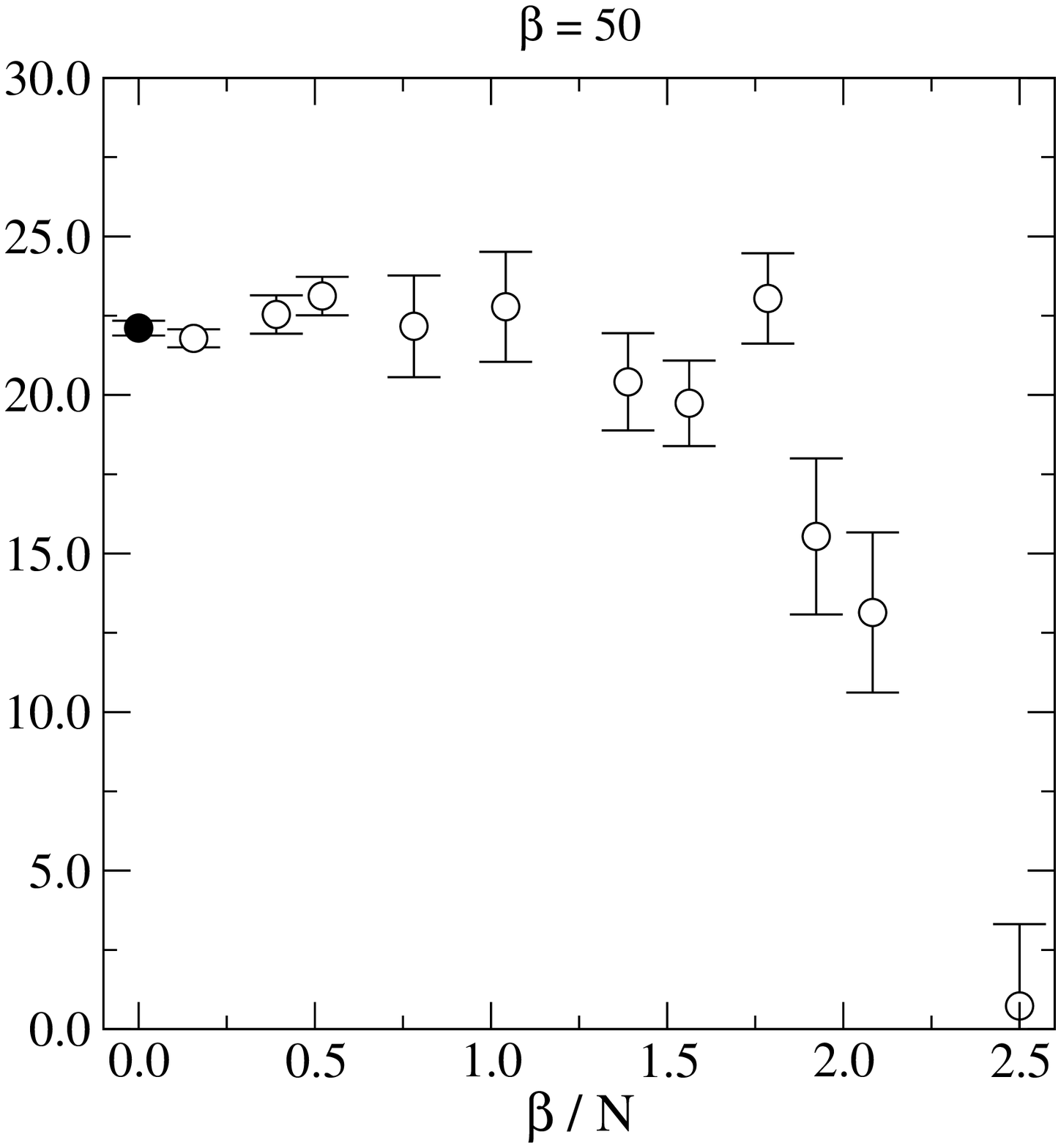}%
}



\caption[a]{Finite-volume values for   
$\beta^4\{\langle 1 - \frac{1}{3} \tr[P_{12}] \rangle_a - 
[c_1/\beta+c_2/\beta^2+c_3/\beta^3+c_4 \ln\beta/\beta^4]\}$, 
as a function of the physical
extent $\beta/N = 6/ g_3^2 L$ of the box. 
The solid symbols indicate 
the infinite-volume estimates, obtained by fitting a constant
to data in the range $\beta/N < 1$.}

\la{fig:infvol}
\end{figure}

Given the infinite-volume estimates, we extrapolate the data to the 
continuum limit, $\beta\to\infty$. In \fig\ref{fig:final} we show the 
functions $\beta^4\{\langle 1 - \frac{1}{3} \tr[P_{12}] \rangle_a - 
[c_1/\beta+c_2/\beta^2+c_3/\beta^3]\}$ and 
$\beta^4\{\langle 1 - \frac{1}{3} \tr[P_{12}] \rangle_a - 
[c_1/\beta+c_2/\beta^2+c_3/\beta^3+c_4 \ln\beta/\beta^4]\}$. It is 
observed how even the 4-loop logarithmic divergence is visible in 
the data, as some upwards 
curvature for $1/\beta \lsim 0.06$. On the other hand, 
for $1/\beta \le 0.01$ the significance loss due to the subtractions grows
rapidly and the error bars become quite large, so that these
data points have little effect on the fit.

\begin{figure}[t]


\centerline{
~~~\epsfysize=7.0cm\epsfbox{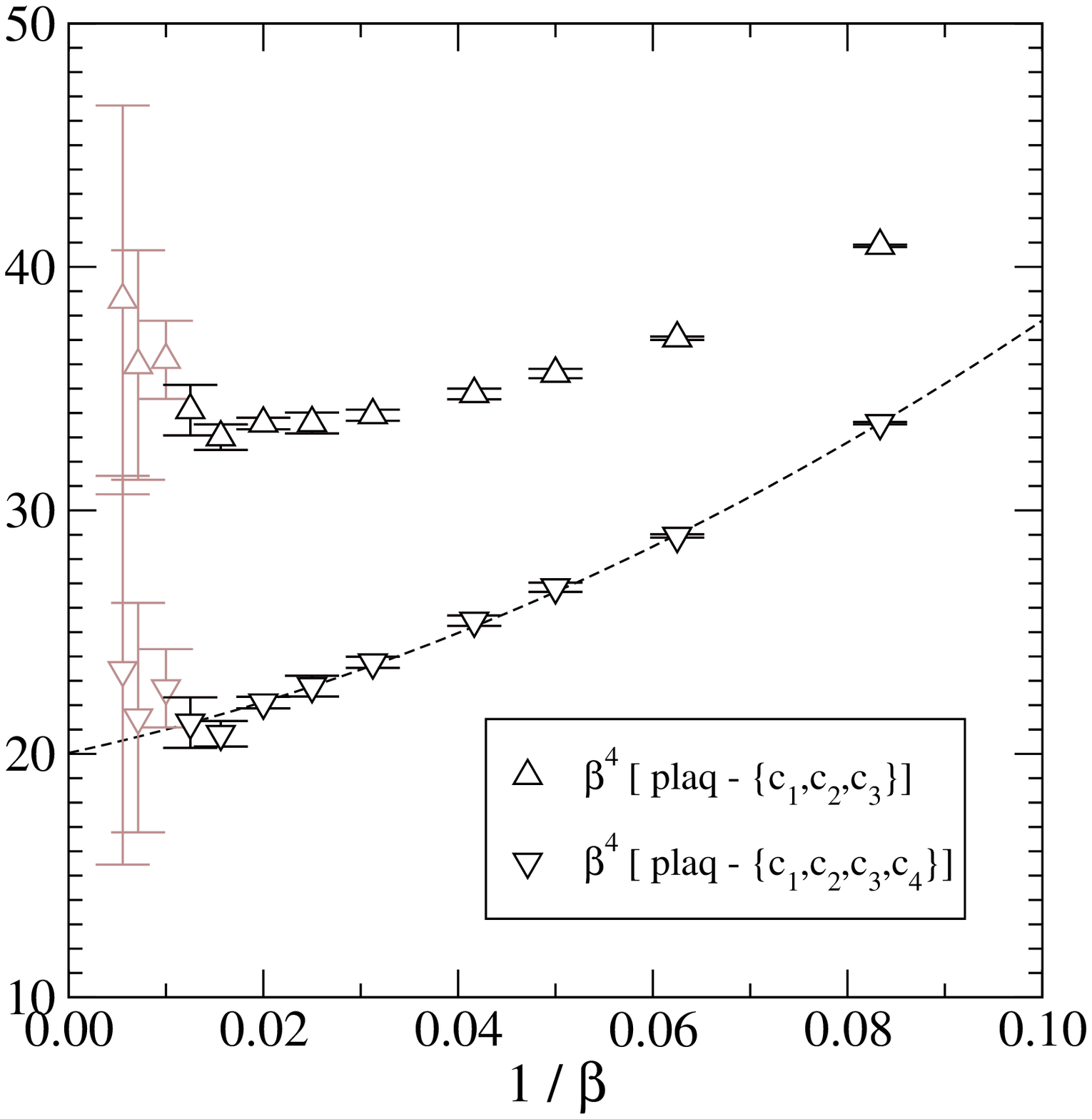}%
}


\caption[a]{The infinite-volume extrapolated data, plotted 
as in \fig\ref{fig:orders}. The effect of the 4-loop logarithmic
divergence is to cause additional upwards ``curvature'' 
in the upper data set. The lower set 
includes all the subtractions, and should thus have a finite continuum limit. 
The continuum extrapolation (as described in the text) is indicated
with the dashed line. The  gray points have error bars so large that 
they are insignificant as far as the fit is concerned.}

\la{fig:final}
\end{figure}

The continuum extrapolation is carried out by fitting a function 
$d_1+d_2/\beta+d_3/\beta^2$ to the infinite-volume extrapolated
data for $\beta^4\{\langle 1 - \frac{1}{3} \tr[P_{12}] \rangle_a - 
[c_1/\beta+c_2/\beta^2+c_3/\beta^3+c_4 \ln\beta/\beta^4]\}$,
in the range $0.01 < 1/\beta < 0.10$. We find that this functional 
form describes the data very well. 
The fitted values are 
  $d_1 = 19.4 ... 20.7$, 
  $d_2 = 110 ... 63$, 
  $d_3 = 717 ... 1101$, 
with $\chi^2/\mbox{dof} = 5.8/6$, where
the intervals indicate the projections of the 68\% confidence
level contour \pagebreak (i.e. the surface 
where $\chi^2 = \chi^2_\rmi{min} + 3.53$) 
onto the various axes, from one end of the elongated ellipse
to the other.\footnote{ 
  If the three largest $\beta$'s are included in the fit, 
  the parameters remain essentially the same, 
  $d_1 = 19.4 ... 20.8$, 
  $d_2 = 107 ... 62$, 
  $d_3 = 733 ... 1117$, 
  while $\chi^2/\mbox{dof} = 7.0/9$
  has decreased due to the large error bars at these $\beta$'s.
  }
We have also estimated the systematic errors from the effect of higher order 
terms in the fit ansatz, and found that they are of the same order 
as these intervals, which we thus consider as our combined error estimates.
Returning back to \eq\nr{eq:betaG}, we then obtain our final result,
\ba
 \bG + 
 \biggl( \frac{43}{12} - \frac{157}{768} \pi^2 \biggr) c_4'
 = \biggl( \frac{2\pi^2}{27} \biggr)^2 \times ( 20.0 \pm 0.7 ) 
 = 10.7 \pm 0.4
 \;, \la{eq:final}
\ea
where we have inserted $\Nc = 3$.

%
\section{Conclusions}
\la{se:conclusions}

The purpose of this paper has been to study the expectation value of
the elementary plaquette in pure SU(3) lattice gauge theory in three
dimensions, as well as to outline how
the $\msbar$ scheme gluon condensate of the 
continuum theory can be extracted from it.  To achieve this goal, 
we have carried out high precision numerical Monte Carlo
simulations close to the continuum limit, corresponding to 
lattice spacings $0.05
\lsim a g_3^2 \lsim 0.5$, where $g_3^2$ is the gauge coupling.

When the leading perturbative terms, up to 4-loop level, are subtracted from 
the plaquette expectation value, and the result is divided by $(a g_3^2)^4$, 
a finite quantity remains (the right-hand side of \eq\nr{eq:betaG},
without $c_4'$) which can be taken as the definition of 
a renormalised gluon condensate in lattice regularization (in certain units).
We have carried out the subtractions and the extrapolation
$a g_3^2 \to 0$, 
and shown that our data 
appear to be precise
enough to determine the remainder with less than 5\% errors, 
cf.\ \fig\ref{fig:final} and \eq\nr{eq:final}. 

To relate this number to the gluon condensate in some continuum scheme, 
say $\msbar$, a further perturbative 4-loop matching computation remains
to be completed, fixing the constant $c_4'$ 
in \eqs\nr{eq:betaG}, \nr{eq:final}. 
Our study should provide a strong incentive for 
finalising this challenging but feasible task, and 
there indeed is work in progress with this goal. 
The $\msbar$ scheme conversion is 
also needed in order to apply our result in the context of finite 
temperature physics, particularly for determining the 
$\mathcal{O}(g^6T^4)$ contribution to the pressure of hot QCD, 
since the other parts of that computation have been formulated 
in the $\msbar$ scheme~\cite{gsixg}.

%
\section*{Acknowledgements}

We are grateful to H.~Panagopoulos and A.~Tsapalis for 
disclosing the results of Ref.~\cite{pt} prior to publication.
This work was partly supported by the Academy
of Finland, contracts no.\ 77744, 80170, and 104382, 
as well as by the Magnus Ehrnrooth Foundation.  
Simulations were carried out at the Finnish 
Center for Scientific Computing (CSC); 
the total amount of computing power used was about $4.5 \times 10^{16}$ flop.

\newpage


\end{document}